\documentclass[aps,preprint,superscriptaddress,preprintnumbers]{revtex4-2}
\usepackage[T1]{fontenc}
\usepackage{times}
\usepackage{graphicx,color}
\usepackage{amsfonts,amsmath,amssymb,amsbsy}
\usepackage[colorlinks=true, linkcolor=blue, citecolor=blue, urlcolor=blue]{hyperref}
\usepackage{float,lettrine}
\renewcommand{\figurename}{\textbf{Figure}}

\def\section#1{\medskip\noindent\textbf{#1}\par}
\let\mathbf=\boldsymbol
\def\blue#1{\textcolor{blue}{#1}}
\def\emph#1{\textcolor{magenta}{#1}}
\makeatletter
\renewcommand{\fnum@figure}{\figurename~\textbf{\thefigure}}
\makeatother
\begin{document}
%%%%%%%%%%%%%%%%%%%%%%%%%%%%%%%%%%%%%%%%%%%%%%%%%%%%%%%%%%%%

\title{{\Large Configurable pixelated skyrmions on nanoscale magnetic grids}}

\author{Xichao Zhang}
%\email[Email:~]{zhangxichao_jsps@shinshu-u.ac.jp}
\thanks{These authors contributed equally.}
\affiliation{Department of Electrical and Computer Engineering, Shinshu University, 4-17-1 Wakasato, Nagano 380-8553, Japan}

\author{Jing Xia}
%\email[Email:~]{jingxia@link.cuhk.edu.cn}
\thanks{These authors contributed equally.}
\affiliation{Department of Electrical and Computer Engineering, Shinshu University, 4-17-1 Wakasato, Nagano 380-8553, Japan}

\author{Keiichiro Shirai}
%\email[Email:~]{keiichi@shinshu-u.ac.jp}
\affiliation{Department of Electrical and Computer Engineering, Shinshu University, 4-17-1 Wakasato, Nagano 380-8553, Japan}

\author{Hiroshi Fujiwara}
%\email[Email:~]{fujiwara@cs.shinshu-u.ac.jp}
\affiliation{Department of Electrical and Computer Engineering, Shinshu University, 4-17-1 Wakasato, Nagano 380-8553, Japan}

\author{Oleg A. Tretiakov}
%\email[Email:~]{o.tretiakov@unsw.edu.au}
\affiliation{School of Physics, The University of New South Wales, Sydney 2052, Australia}

\author{\\ Motohiko Ezawa}
\email[Email:~]{ezawa@ap.t.u-tokyo.ac.jp}
\affiliation{Department of Applied Physics, The University of Tokyo, 7-3-1 Hongo, Tokyo 113-8656, Japan}

\author{Yan Zhou}
\email[Email:~]{zhouyan@cuhk.edu.cn}
\affiliation{School of Science and Engineering, The Chinese University of Hong Kong, Shenzhen, Guangdong 518172, China}

\author{Xiaoxi Liu}
\email[Email:~]{liu@cs.shinshu-u.ac.jp}
\affiliation{Department of Electrical and Computer Engineering, Shinshu University, 4-17-1 Wakasato, Nagano 380-8553, Japan}

%-%-%-%-%-%-%-%-%-%-%-%-%-%-%-%-%-%-%-%-%-%-%-%-%-%-%-%-%-%-%
{\sffamily
\begin{abstract}
\noindent
Topological spin textures can serve as non-volatile information carriers. Here we study the current-induced dynamics of an isolated magnetic skyrmion on a nanoscale square-grid pinning pattern formed by orthogonal defect lines with reduced magnetic anisotropy. The skyrmion on the square grid can be pixelated with a quantized size of the grid. We demonstrate that the position, size, and shape of skyrmion on the square grid are electrically configurable. The skyrmion center is quantized to be on the grid and the skyrmion may show a hopping motion instead of a continuous motion. We find that the skyrmion Hall effect can be perfectly prohibited due to the pinning effect of the grid. The pixelated skyrmion can be harnessed to build future programmable racetrack memory, multistate memory, and logic computing device. Our results will be a basis for digital information storage and computation based on pixelated topological spin textures on artificial pinning patterns.
\end{abstract}}
%-%-%-%-%-%-%-%-%-%-%-%-%-%-%-%-%-%-%-%-%-%-%-%-%-%-%-%-%-%-%

\date{\today}

\preprint{\href{https://doi.org/10.1038/s42005-021-00761-7}{\blue{Commun. Phys. \textbf{4}, 255 (2021). DOI:~10.1038/s42005-021-00761-7}}}


\maketitle

\clearpage

%-%-%-%-%-%-%-%-%-%-%-%-%-%-%-%-%-%-%-%-%-%-%-%-%-%-%-%-%-%-%
\section{\sffamily Introduction}
%-%-%-%-%-%-%-%-%-%-%-%-%-%-%-%-%-%-%-%-%-%-%-%-%-%-%-%-%-%-%

\noindent
Topological spin textures in magnets are versatile objects that can be used to facilitate many magnetic and spintronic applications~\cite{Nagaosa_NN2013,Mochizuki_JPCM2015,Wiesendanger_NRM2016,Finocchio_JPD2016,Kang_PIEEE2016,Kanazawa_AM2017,Wanjun_PR2017,Fert_NRM2017,Everschor_JAP2018,Zhou_NSR2018,Diep_2019,Zhang_JPCM2020,Gobel_PR2020,Back_JPD2020,Bogdanov_NPR2020,Li_MH2021,Luo_APL2021,Reichhardt_2021}.
The basic topological spin textures in magnets with Dzyaloshinskii-Moriya (DM) interactions are called magnetic skyrmions~\cite{Roszler_NATURE2006}, which have been created experimentally~\cite{Muhlbauer_SCIENCE2009,Yu_NATURE2010,Heinze_NP2011,Schulz_NP2012,Romming_SCIENCE2013,Wanjun_SCIENCE2015,Woo_NM2016,ML_NN2016,Boulle_NN2016,Soumyanarayanan_NM2017,Mandru_NC2020,Zeissler_NC2020,Birch_NC2021}
and suggested for use in spintronic functional devices, including memories~\cite{Fert_NN2013,Sampaio_NN2013,Tomasello_SREP2014,Koshibae_JJAP2015,Yu_NL2016,Xichao_NC2016,Xichao_PRB2016B,Tomasello_JPD2017,Wang_NC2020},
logic computing gates~\cite{Xichao_SREP2015B,Luo_NL2018,Chauwin_PRA2019,Mankalale_TED2019,Walker_APL2021,Liang_2019},
field-effect transistors~\cite{Xichao_SREP2015A,Upadhyaya_PRB2015,Hong_APL2019},
memristors~\cite{Luo_EDL2019,Jiang_APL2019},
and other novel applications~\cite{Huang_N2017,Li_N2017,Pinna_PRA2018,Prychynenko_PRA2018,Nozaki_APL2019,Zazvorka_NN2019,Song_NE2020,Jibiki_APL2020}.

Toward the realization of device applications based on skyrmions, a number of theoretical and experimental works have been devoted to the studies of statics and dynamics of skyrmions~\cite{Nagaosa_NN2013,Mochizuki_JPCM2015,Wiesendanger_NRM2016,Finocchio_JPD2016,Kang_PIEEE2016,Kanazawa_AM2017,Wanjun_PR2017,Fert_NRM2017,Everschor_JAP2018,Zhou_NSR2018,Diep_2019,Zhang_JPCM2020,Gobel_PR2020,Back_JPD2020,Bogdanov_NPR2020,Li_MH2021,Luo_APL2021,Reichhardt_2021}.
In particular, the skyrmion texture is found to have several degrees of freedom that, in principle, could carry useful information and could be controlled and manipulated independently.
For example, the position of a skyrmion can be manipulated by a spin current~\cite{Wanjun_SCIENCE2015,Woo_NM2016,Zeissler_NC2020,Fert_NN2013,Sampaio_NN2013,Tomasello_SREP2014,Koshibae_JJAP2015,Yu_NL2016,Xichao_NC2016,Xichao_PRB2016B,Tomasello_JPD2017,Xichao_N2015,Liu_PRA2019,Ma_NL2019,Wanjun_NPHYS2017,Litzius_NPHYS2017,Dohi_NC2019};
the size of a skyrmion can be controlled by an out-of-plane magnetic field~\cite{Rohart_PRB2013,Woo_NM2016,ML_NN2016,Boulle_NN2016};
the topological charge of a skyrmion can be changed by a vertical spin current~\cite{Xichao_PRB2013};
the polarity of a skyrmion can be switched by magnetic fields and spin currents~\cite{Liu_PRB2015,Beg_SREP2015,Zhang_APL2015,Zheng_PRL2017};
the helicity of a skyrmion in frustrated magnets can be manipulated by a spin current~\cite{Lin_PRB2016A,Xichao_NCOMMS2017,Xia_PRApplied2019,Zhang_PRB2020,Xia_APL2020A,Xichao_APL2021}.

Typically, an isolated skyrmion stabilized by the DM interaction shows a circular shape~\cite{Roszler_NATURE2006,Rohart_PRB2013}.
Recently, skyrmions with different shapes have been realized~\cite{Lin_PRB2014,Khanh_NN2020,Hayami_PRB2021,Peng_NN2020,Xia_APL2020B,Jena_NC2020,Cui_AM2021,Ma_PRB2016}.
For example, a square lattice of square-shaped skyrmions was discovered by Khanh \textit{et al.}~\cite{Khanh_NN2020} in a centrosymmetric tetragonal magnet, of which the origin was theoretically investigated by Hayami and Motome~\cite{Hayami_PRB2021}.
A square lattice of square-shaped antiskyrmions was also observed experimentally by Peng \textit{et al.}~\cite{Peng_NN2020} in a noncentrosymmetric magnet.
Besides, it is found that skyrmions and antiskyrmions can show elliptical shapes in samples with anisotropic DM interactions~\cite{Xia_APL2020B,Jena_NC2020,Cui_AM2021}.
The deformation of a skyrmion induced by external forces may also result in a non-circular shape~\cite{Zeissler_NC2020,Litzius_NPHYS2017}.
All these findings on skyrmions showing different shapes reflect the importance of controlling the shape of a skyrmion, which may lead to novel spintronic applications based on topological spin textures with different shapes.

Recently, Juge \textit{et al.}~\cite{Juge_NL2021} and Ohara \textit{et al.}~\cite{Ohara_NL2021} independently demonstrated the control of skyrmion position by locally modifying the magnetic properties.
They experimentally realized the confinement of skyrmions in nanoscale tracks with modified magnetic properties on a large film.
In particular, the local modification of perpendicular magnetic anisotropy (PMA) can result in an energy barrier, which plays a key role on the confinement and pinning of skyrmions.
Therefore, by locally modifying PMA or other magnetic properties it is envisioned that one can fabricate different types of artificial pinning patterns on magnetic materials, such as parallel defect lines, grids, and square patterns~\cite{Reichhardt_PRB2010,Reichhardt_PRL2015,Reichhardt_PRB2015A,Reichhardt_PRB2015B,Reichhardt_NJP2015,Reichhardt_NJP2016,Reichhardt_PRB2016,Ma_PRB2016,Muller_NJP2017,Reichhardt_PRB2018,Fernandes_NC2018,Chen_PRA2020,Navau_PRB2019,Navau_NANOSCALE2019,Bhatti_2019,Reichhardt_PRB2020,Vizarim_PRB2020,Feilhauer_PRB2020,Lounis_JPCM2020,Lounis_SR2020,Menezes_PRB2019,Vizarim_JPCM2021,Reichhardt_2021}.
These artificial pinning patterns may lead to very special static and dynamic behaviors of topological spin textures interacting with them~\cite{Reichhardt_PRB2010,Reichhardt_PRL2015,Reichhardt_PRB2015A,Reichhardt_PRB2015B,Reichhardt_NJP2015,Reichhardt_NJP2016,Reichhardt_PRB2016,Ma_PRB2016,Muller_NJP2017,Reichhardt_PRB2018,Fernandes_NC2018,Chen_PRA2020,Navau_PRB2019,Navau_NANOSCALE2019,Bhatti_2019,Reichhardt_PRB2020,Vizarim_PRB2020,Feilhauer_PRB2020,Lounis_JPCM2020,Lounis_SR2020,Menezes_PRB2019,Vizarim_JPCM2021,Reichhardt_2021}.
For example, the skyrmion Hall effect can be controlled or reduced for skyrmions moving over two-dimensional periodic pinning arrays in certain cases~\cite{Reichhardt_PRB2015A,Feilhauer_PRB2020}.
Moreover, the artificial pinning patterns would also offer the possibility to study basic science issues since particles and quasi-particles (e.g., superconducting vortices and colloids) on periodic substrates is a wide-ranging problem~\cite{Coppersmith_1982,Harada_1996,Martin_1997,Reichhardt_1998,Berdiyorov_2006,Tung_2006,Bohlein_2012,Reichhardt_2021}.

In this work, we report the properties of a skyrmion in a magnetic thin film with the square-grid pinning pattern formed by nanoscale orthogonal defect lines with reduced PMA.
We show that the square grid leads to the formation of pixelated skyrmions, of which the position and area are quantized in the unit of the grid cell.
We find that the position, size, and shape of a pixelated skyrmion can be manipulated precisely by a current pulse.

%-%-%-%-%-%-%-%-%-%-%-%-%-%-%-%-%-%-%-%-%-%-%-%-%-%-%-%-%-%-%
\vbox{}
\section{\sffamily Results and Discussion}
%-%-%-%-%-%-%-%-%-%-%-%-%-%-%-%-%-%-%-%-%-%-%-%-%-%-%-%-%-%-%

\begin{figure}[h]
\centerline{\includegraphics[width=0.90\textwidth]{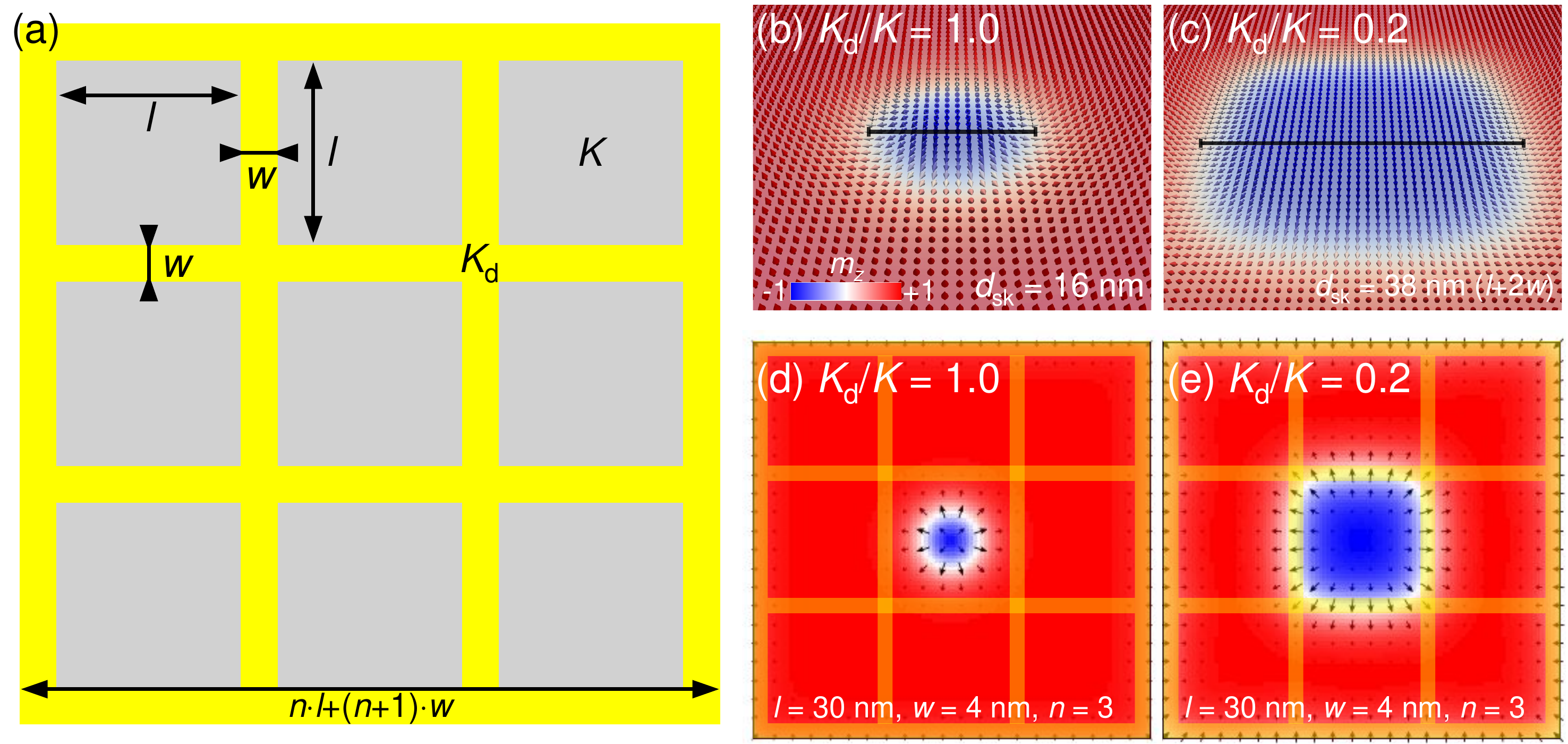}}
\caption{%
\textbf{An ordinary magnetic skyrmion and a square-shaped magnetic skyrmion.}
(\textbf{a}) Top-view schematic of the simulation geometry. $l$ denotes the spacing between two adjacent parallel defect lines. $w$ denotes the width of the defect line. $n$ denotes the number of unit square patterns along the $x$ and $y$ directions. The total side length of the ferromagnetic layer equals $n\cdot l+(n+1)\cdot w$. $K$ and $K_{\text{d}}$ stand for the perpendicular magnetic anisotropy constants for the unmodified areas and defect lines, respectively. The defect lines with reduced $K_{\text{d}}$ are indicated by yellow lines. The thickness of the ferromagnetic layer is fixed at $1$ nm.
(\textbf{b}) Illustration of a relaxed ordinary round-shaped skyrmion with the topological charge of $Q=1$ at the center of a sample with $K_{\text{d}}/K=1$. The side length of the sample equals $106$ nm. The relaxed skyrmion diameter $d_{\text{sk}}$ equals $16$ nm, which is indicated by the black line. The color scale represents the out-of-plane magnetization component $m_z$, which has been used throughout the work.
(\textbf{c}) Illustration of a relaxed square-shaped skyrmion with $Q=1$ at the center of a sample with $K_{\text{d}}/K=0.2$. Here, $l=30$ nm, $w=4$ nm, and $n=3$. The side length of the sample equals $106$ nm. The relaxed skyrmion diameter $d_{\text{sk}}$ equals $38$ nm, as indicated by the black line.
(\textbf{d}) Top view of the sample with $K_{\text{d}}/K=1$, corresponding to (\textbf{b}).
(\textbf{e}) Top view of the sample with $K_{\text{d}}/K=0.2$, corresponding to (\textbf{c}).
}
\label{FIG1}
\end{figure}

\noindent
\textbf{Static properties of the square-shaped skyrmion.}
Figure~\ref{FIG1}(a) depicts the simulation geometry.
We consider a ferromagnetic (FM) thin layer attached to a heavy-metal layer, where the FM layer has certain PMA and interface-induced DM interaction. The FM layer thickness is fixed at $1$ nm in all simulations.
We assume that the square-grid pinning pattern in the FM layer is formed by orthogonal defect lines with reduced PMA, which can be realized in experiments by locally modifying the magnetic properties (\textit{i.e.}, using additional sputtered layers or ion irradiation)~\cite{Ohara_NL2021,Juge_NL2021,Jong_2021}.
The width of each defect line is defined as $w$. The distance between two nearest-neighboring parallel defect lines is defined as $l$. The number of unit square patterns along the $x$ or $y$ directions is defined as $n$. Hence, the total side length of the FM layer is equal to $nl+(n+1)w$ [see Fig.~\ref{FIG1}(a)].
The magnetic parameters and other modeling details are given in the \blue{Methods}.

We first study a static single isolated skyrmion in the sample with the square grid.
At the initial state, a N{\'e}el-type skyrmion with a theoretical topological charge $Q=1$ is placed at the sample center, which is relaxed to a stable or metastable state by the OOMMF conjugate gradient minimizer~\cite{OOMMF}.
The topological charge $Q$ is defined as
$Q=-\frac{1}{4\pi}\int\boldsymbol{m}\cdot(\frac{\partial\boldsymbol{m}}{\partial x}\times\frac{\partial\boldsymbol{m}}{\partial y})dxdy$
with $\boldsymbol{m}$ being the reduced magnetization.
In this work, we assume that the initial skyrmion diameter is smaller than the defect-line spacing $l$. For example, the relaxed ordinary skyrmion has a diameter $d_{\text{sk}}$ of $16$ nm in the sample with $l=30$ nm, $w=4$ nm, $n=3$, and $K_{\text{d}}/K=1$ [see Fig.~\ref{FIG1}(b)]. $K_{\text{d}}/K=1$ means no square-grid pinning pattern exists in the sample.
However, when $K_{\text{d}}/K<1$, the initial skyrmion may relax to a square-shaped skyrmion with an enlarged size determined by the square-grid pinning pattern, as shown in Fig.~\ref{FIG1}(c). In particular, the diameter (\textit{i.e.}, the side length) of the square-shaped skyrmion $d_{\text{sk}}$ is found to be $l+2w$. For example, in Fig.~\ref{FIG1}(c) the relaxed square-shaped skyrmion shows $d_{\text{sk}}=38$ nm in the sample with $l=30$ nm, $w=4$ nm, $n=3$, and $K_{\text{d}}/K=0.2$.
We note that the square-shaped skyrmion with a larger size will be easier to be detected and observed.
Once a square-shaped skyrmion is formed on the grid, its position and area are determined in the unit of the grid cell. As a minute grid cell is regarded as a pixel of the grid, the square-shaped skyrmion on the grid is seen as a pixelated skyrmion with quantized area.

Figures~\ref{FIG1}(d) and~\ref{FIG1}(e) show the top views of an ordinary skyrmion at $K_{\text{d}}/K=1$ and a square-shaped skyrmion at $K_{\text{d}}/K=0.2$, respectively.
The formation of the square-shaped skyrmion on the grid with $K_{\text{d}}/K<1$ is due to the fact that the defect line with reduced PMA attracts the domain wall or skyrmion near it~\cite{Ohara_NL2021}.
The formation of the square-shaped skyrmion is favored by smaller $K_{\text{d}}/K$ and wider defect lines provided that $0\leq K_{\text{d}}/K\leq 1$ and $w\ll d_{\text{sk}}<l$ (see \blue{Supplementary Note 1}).
We note that the square-shaped skyrmion is formed on the grid in the absence of an out-of-plane magnetic field $B_z$. However, the size of the square-shaped skyrmion is adjusted by $B_z$ when $K_{\text{d}}$ is much smaller than $K$ (see \blue{Supplementary Note 2}).

\begin{figure}[t]
\centerline{\includegraphics[width=0.90\textwidth]{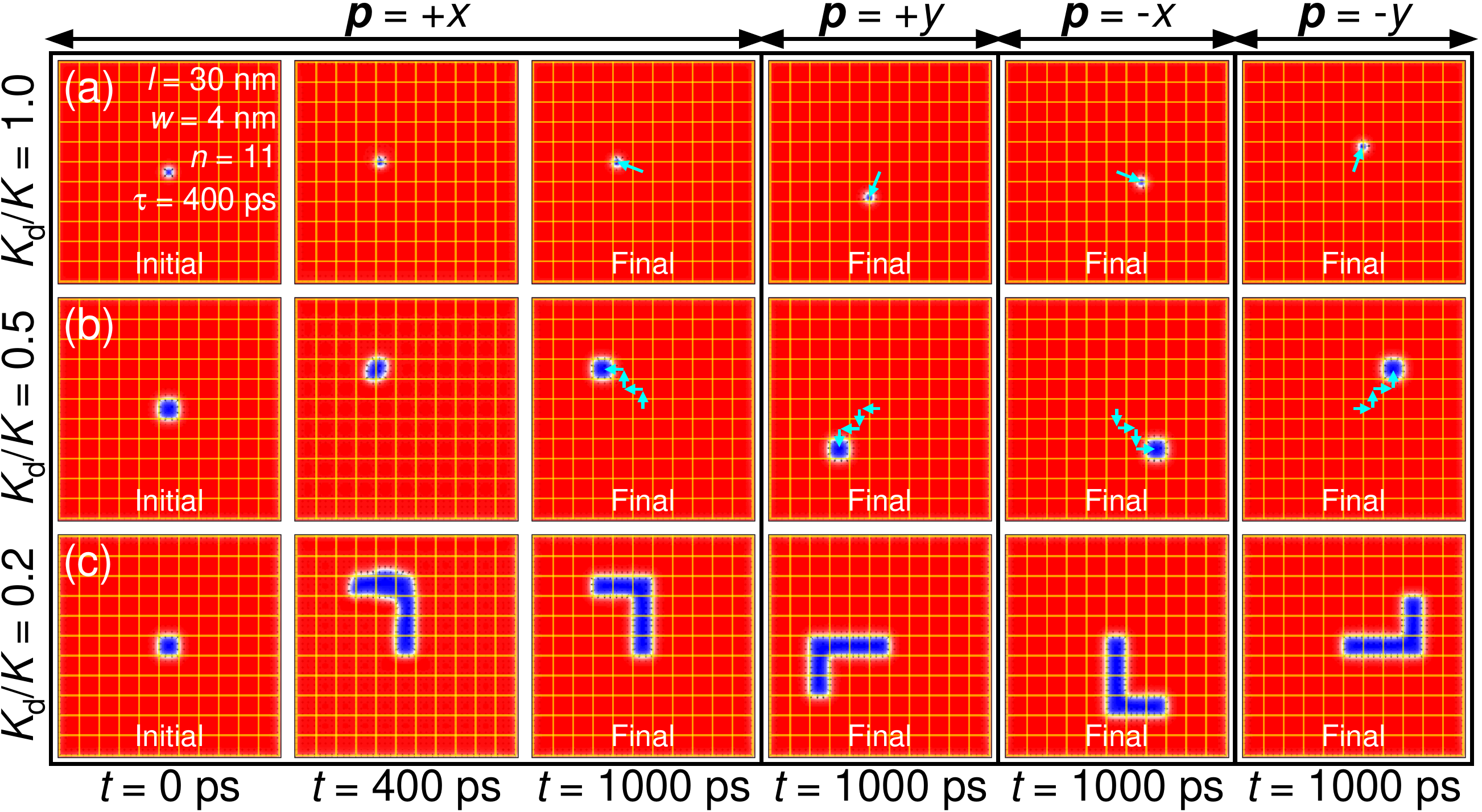}}
\caption{%
\textbf{Current-induced dynamics of a square-shaped skyrmion on the grids.}
(\textbf{a}) Top views of the current-induced motion of an ordinary round-shaped skyrmion in a sample with $K_{\text{d}}/K=1$ for different spin polarization direction $\boldsymbol{p}$. $K$ and $K_{\text{d}}$ stand for the perpendicular magnetic anisotropy constants for the unmodified areas and defect lines, respectively. The side length of the sample equals $378$ nm. An ordinary skyrmion with the topological charge of $Q=1$ is relaxed at the sample center as the initial state at $t=0$ ps. A $400$-ps-long current pulse of $j=100$ MA cm$^{-2}$ is applied, and then the system is relaxed for $600$ ps. Here the damping parameter $\alpha=0.3$.
(\textbf{b}) Top views of the current-induced hopping motion of a square-shaped skyrmion in a sample with defect lines for different $\boldsymbol{p}$. Here, $l=30$ nm, $w=4$ nm, $n=11$, and $K_{\text{d}}/K=0.5$. $l$ denotes the spacing between two adjacent parallel defect lines. $w$ denotes the width of the defect line. $n$ denotes the number of unit square patterns along the $x$ and $y$ directions. The side length of the sample equals $378$ nm. A square-shaped skyrmion with $Q=1$ is relaxed at the sample center as the initial state at $t=0$ ps. A $400$-ps-long current pulse of $j=100$ MA cm$^{-2}$ is applied, and then the system is relaxed for $600$ ps. The defect lines with reduced $K_{\text{d}}$ are indicated by yellow lines. The motion route is denoted by the cyan arrow.
(\textbf{c}) Top views of the current-induced deformation of a square-shaped skyrmion in a sample with defect lines for different $\boldsymbol{p}$. Here $K_{\text{d}}/K=0.2$. The other parameters are the same as those in (\textbf{b}).
}
\label{FIG2}
\end{figure}

\vbox{}
\noindent
\textbf{Current-induced dynamics of the square-shaped skyrmion.}
We also study the current-induced dynamics of a square-shaped skyrmion on the grid (see Fig.~\ref{FIG2}).
We consider a sample at $B_z=0$ mT with $l=30$ nm, $w=4$ nm, and $n=11$. The total side length of the sample is equal to $378$ nm.
At the initial state (\textit{i.e.}, $t=0$ ps), we place a relaxed skyrmion with $Q=1$ at the sample center.
We first apply a single current pulse to drive the dynamics at a damping parameter of $\alpha=0.3$. The pulse length is fixed at $\tau=400$ ps. The current density is fixed at $j=100$ MA cm$^{-2}$. After the application of the pulse, the sample is relaxed for $600$ ps. The total simulation time equals $1000$ ps.
Note that we only consider the damping-like spin-orbit torque generated by the current pulse (see \blue{Methods}).
The field-like torque contribution in our material system could be very small compared to the damping-like torque as our system does not have a large interfacial Rashba effect. Also, a large field-like torque usually leads to the deformation of a skyrmion~\cite{Litzius_NPHYS2017}, which cannot provide a driving force.

As shown in Fig.~\ref{FIG2}(a), an ordinary skyrmion shows directional motion in the sample with $K_{\text{d}}/K=1$ when a single current pulse is applied.
The direction of motion depends on the spin polarization direction $\boldsymbol{p}$, which is controlled by the current injection direction in experiments.
When $\boldsymbol{p}=+\hat{x}$, the ordinary skyrmion moves smoothly toward the $+y$ direction and shows an obvious transverse shift in the $-x$ direction due to the skyrmion Hall effect~\cite{Wanjun_NPHYS2017,Litzius_NPHYS2017}.
The skyrmion stops when the current pulse is off at $t=\tau=400$ ps, and the final state obtained at $t=1000$ ps shows that the skyrmion is closer to the upper left corner of the sample (see \blue{Supplementary Movie 1}).
The skyrmion moves closer to the lower left, lower right, and upper right corners driven by the current pulses with $\boldsymbol{p}=+\hat{y}$, $\boldsymbol{p}=-\hat{x}$, and $\boldsymbol{p}=-\hat{y}$, respectively.

When $K_{\text{d}}/K<1$, we find that the square-shaped skyrmion shows very different current-induced dynamic behaviors, which depend on the value of $K_{\text{d}}/K$.
In the sample with $K_{\text{d}}/K=0.5$, the square-shaped skyrmion shows a hopping motion instead of a smooth motion [see Fig.~\ref{FIG2}(b)], which is caused by the pinning effect of the grid (see \blue{Supplementary Movie 2})
The hopping motion toward the sample corner is a result of the skyrmion Hall effect.
However, in the sample with $K_{\text{d}}/K=0.2$, the square-shaped skyrmion shows directional elongation and deformation when the current pulse is applied [see Fig.~\ref{FIG2}(c)].
Due to the skyrmion Hall effect, the square-shaped skyrmion is transformed to a pixelated L-shaped skyrmion (see \blue{Supplementary Movie 3}), which carries a theoretical topological charge of $Q=1$.
We find that depending on $\boldsymbol{p}$, the position and orientation of the final L-shaped skyrmion can be controlled.

\begin{figure}[t]
\centerline{\includegraphics[width=0.90\textwidth]{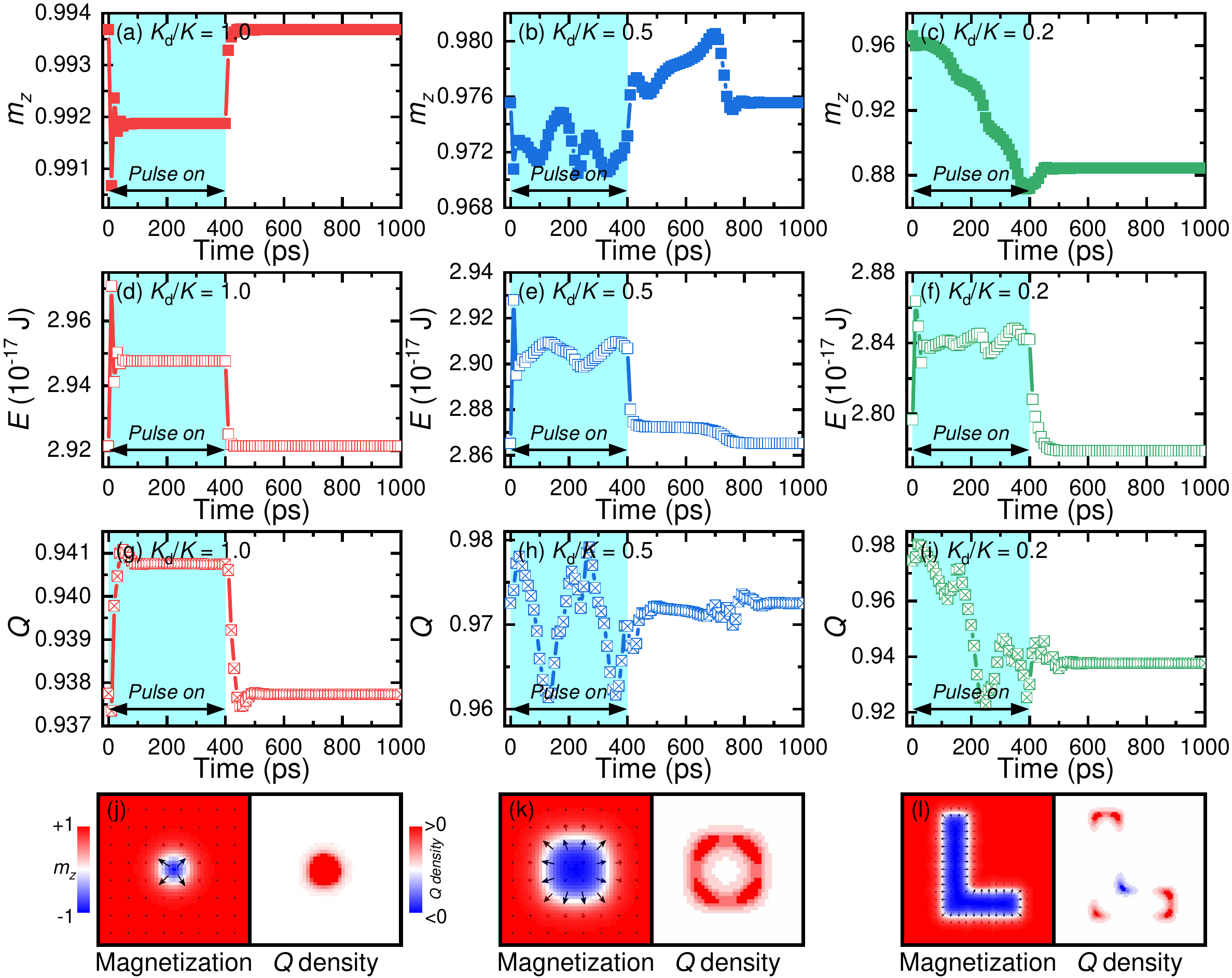}}
\caption{%
\textbf{Out-of-plane magnetization, energy, and topological charge of the current-driven square-shaped skyrmion.}
(\textbf{a}) Time-dependent out-of-plane magnetization $m_z$ corresponding to the current-induced motion of an ordinary round-shaped skyrmion in a sample with $K_{\text{d}}/K=1$. $K$ and $K_{\text{d}}$ stand for the perpendicular magnetic anisotropy constants for the unmodified areas and defect lines, respectively. A $400$-ps-long current pulse of $j=100$ MA cm$^{-2}$ and $\boldsymbol{p}=+\hat{x}$ is applied, and then the system is relaxed for $600$ ps.
(\textbf{b}) Time-dependent $m_z$ corresponding to the current-induced hopping motion of a square-shaped skyrmion in a sample with defect lines of $K_{\text{d}}/K=0.5$.
(\textbf{c}) Time-dependent $m_z$ corresponding to the current-induced deformation of a square-shaped skyrmion in a sample with defect lines of $K_{\text{d}}/K=0.2$.
(\textbf{d}) Time-dependent total energy $E$ of the system corresponding to (\textbf{a}).
(\textbf{e}) Time-dependent $E$ corresponding to (\textbf{b}).
(\textbf{f}) Time-dependent $E$ corresponding to (\textbf{c}).
(\textbf{g}) Time-dependent numerical topological charge $Q$ of the system corresponding to (\textbf{a}).
(\textbf{h}) Time-dependent numerical $Q$ corresponding to (\textbf{b}).
(\textbf{i}) Time-dependent numerical $Q$ corresponding to (\textbf{c}).
(\textbf{j}) $Q$ density distribution of an ordinary skyrmion.
(\textbf{k}) $Q$ density distribution of a square-shaped skyrmion.
(\textbf{l}) $Q$ density distribution of an L-shaped skyrmion at the final state.
}
\label{FIG3}
\end{figure}

First, we review the results on an ordinary skyrmion with $K_{\text{d}}/K=1$.
$m_z$ slightly decreases to a stable value during the application of the current pulse, indicating the steady motion of the skyrmion with a slightly reduced size during the pulse application as shown in Fig.~\ref{FIG3}(a).
The time-dependent total energy $E$ is given in Fig.~\ref{FIG3}(d), which rapidly recovers to the initial-state value after the pulse application, suggesting that the initial and final states are the same.
Figure~\ref{FIG3}(g) shows the time-dependent numerical topological charge $Q$, which slightly increases during the skyrmion motion.
We note that the numerical $Q$ is not exactly equal to $1$ at the initial and final state. The non-integer value is caused by the discretized meshes and current-induced deformation of the skyrmion. We have excluded the effect of tilted edge spins on the calculation of $Q$.

\begin{figure}[t]
\centerline{\includegraphics[width=0.60\textwidth]{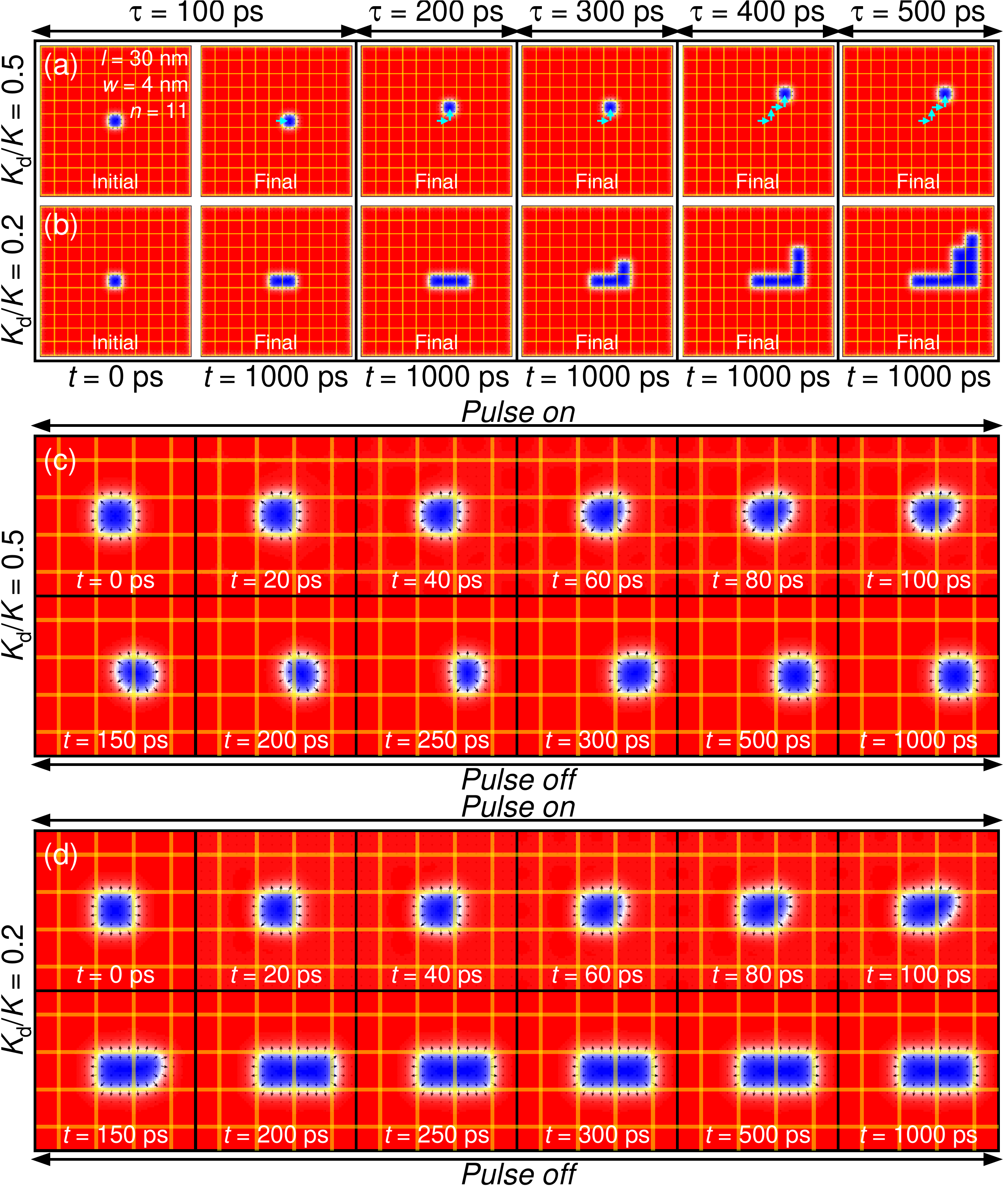}}
\caption{%
\textbf{Effect of pulse length on the current-induced dynamics of a square-shaped skyrmion.}
(\textbf{a}) Top views of the hopping motion of a square-shaped skyrmion driven by a current pulse with the pulse length $\tau$. Here, $l=30$ nm, $w=4$ nm, $n=11$, $K_{\text{d}}/K=0.5$, and the damping parameter $\alpha=0.3$. $K$ and $K_{\text{d}}$ stand for the perpendicular magnetic anisotropy constants for the unmodified areas and defect lines, respectively. $l$ denotes the spacing between two adjacent parallel defect lines. $w$ denotes the width of the defect line. $n$ denotes the number of unit square patterns along the $x$ and $y$ directions. The side length of the sample equals $378$ nm. A square-shaped skyrmion with the topological charge of $Q=1$ is relaxed at the sample center as the initial state at $t=0$ ps. A current pulse of $j=100$ MA cm$^{-2}$ and $\boldsymbol{p}=-\hat{y}$ is applied, and then the system is relaxed until $t=1000$ ps. The defect lines with reduced $K_{\text{d}}$ are indicated by yellow lines. The motion route is denoted by the cyan arrow.
(\textbf{b}) Top views of the deformation of a square-shaped skyrmion driven by a current pulse with $\tau$. Here $K_{\text{d}}/K=0.2$. The other parameters are the same as those in (\textbf{a}).
(\textbf{c}) Zoomed top views of the hopping motion of a square-shaped skyrmion driven by a $100$-ps-long pulse of $j=100$ MA cm$^{-2}$. The other parameters are the same as those in (\textbf{a}).
(\textbf{d}) Zoomed top views of the deformation of a square-shaped skyrmion driven by a $100$-ps-long pulse of $j=100$ MA cm$^{-2}$. The other parameters are the same as those in (\textbf{b}).
}
\label{FIG4}
\end{figure}

Next, we study the case with $K_{\text{d}}/K=0.5$, where a square-shaped skyrmion shows a hopping motion.
$m_z$ oscillates during the pulse application, which implies the oscillating changes of the skyrmion size and shape during its hops across square grid cells as shown in Fig.~\ref{FIG3}(b).
The detailed process is shown in Fig.~\ref{FIG4}(c).
After the pulse application, both $m_z$ and $E$ are recovered to their initial-state values at $t=0$ ps [see Fig.~\ref{FIG3}(e)], justifying that the initial and final skyrmion states are the same because of the translational symmetry of the square-shaped skyrmion.
We note that the numerical $Q$ oscillates during the pulse application [see Fig.~\ref{FIG3}(h)], which is caused by the oscillation of the square-shaped skyrmion.

Finally, we study the case with $K_{\text{d}}/K=0.2$.
As the square-shaped skyrmion is enlarged and deformed during the pulse application, $m_z$ significantly decreases during $t=0-400$ ps and reaches a stable value at $t=500$ ps. See Fig.~\ref{FIG3}(c).
The detailed process is shown in Fig.~\ref{FIG4}(d).
$E$ increases and oscillates during the deformation [see Fig.~\ref{FIG3}(f)], indicating the deformation is produced by overcoming periodic energy barriers on the grid.
The numerical $Q$ oscillates and slightly decreases [see Fig.~\ref{FIG3}(i)], which result from the different sizes and shapes of the initial and final states.
The distributions of local $Q$ density for the ordinary, square-shaped, and L-shaped skyrmions are given in Figs.~\ref{FIG3}(j),~\ref{FIG3}(k), and~\ref{FIG3}(l), respectively. We find that the local $Q$ density is localized at the corners of pixelated skyrmions. Especially, the local $Q$ density is negative for the $270$ degree corner of the L-shaped skyrmion.

In Fig.~\ref{FIG4}, we continue to investigate the effect of the pulse length $\tau$ on the dynamics of a square-shaped skyrmion.
At $t=0$ ps, we place a relaxed square-shaped skyrmion with $Q=1$ at the sample center.
We focus on the dynamics induced by a single current pulse with $j=100$ MA cm$^{-2}$ and $\boldsymbol{p}=-\hat{y}$.
As shown in Fig.~\ref{FIG4}(a), the square-shaped skyrmion in the sample with $K_{\text{d}}/K=0.5$ hops from the grid cell at the sample center to the right nearest-neighboring grid cell after the pulse application with $\tau=100$ ps (see \blue{Supplementary Movie 4}).
It shows no transverse shift because that the applied current pulse is too short to activate the skyrmion Hall effect, as shown in Fig.~\ref{FIG4}(c).
Therefore, by applying a sequence of $100$-ps-long pulses with a pulse spacing of $900$ ps, it is possible to drive the square-shaped skyrmion moves exactly toward the $+x$ direction without showing the skyrmion Hall effect (see \blue{Supplementary Movie 5}).
When $\tau=200-500$ ps, the square-shaped skyrmion shows the skyrmion Hall effect during its hopping motion (see \blue{Supplementary Movie 6}).

In Fig.~\ref{FIG4}(b), the square-shaped skyrmion in the sample with $K_{\text{d}}/K=0.2$ elongates to a rectangle-shaped skyrmion after the pulse application with $\tau=100-200$ ps (see \blue{Supplementary Movie 7}).
As mentioned above, a part of the domain wall forming the skyrmion is pinned by the defect line. Hence, the square-shaped skyrmion is forced to deform along the $+x$ direction when the skyrmion Hall effect is prohibited [see Fig.~\ref{FIG4}(d)].
However, under the pulse application with $\tau=300-400$ ps, the depinned domain wall tends to propagate in both the $+x$ and $+y$ directions due to the skyrmion Hall effect; therefore, the square-shaped skyrmion is deformed into an L-shaped skyrmion (see \blue{Supplementary Movie 8}), which is regarded as an L-shaped skyrmion as it carries a theoretical $Q=1$.
When $\tau=500$ ps, the deformation is more remarkable and complex, leading to an abnormal L-shaped skyrmion.

\begin{figure}[t]
\centerline{\includegraphics[width=0.80\textwidth]{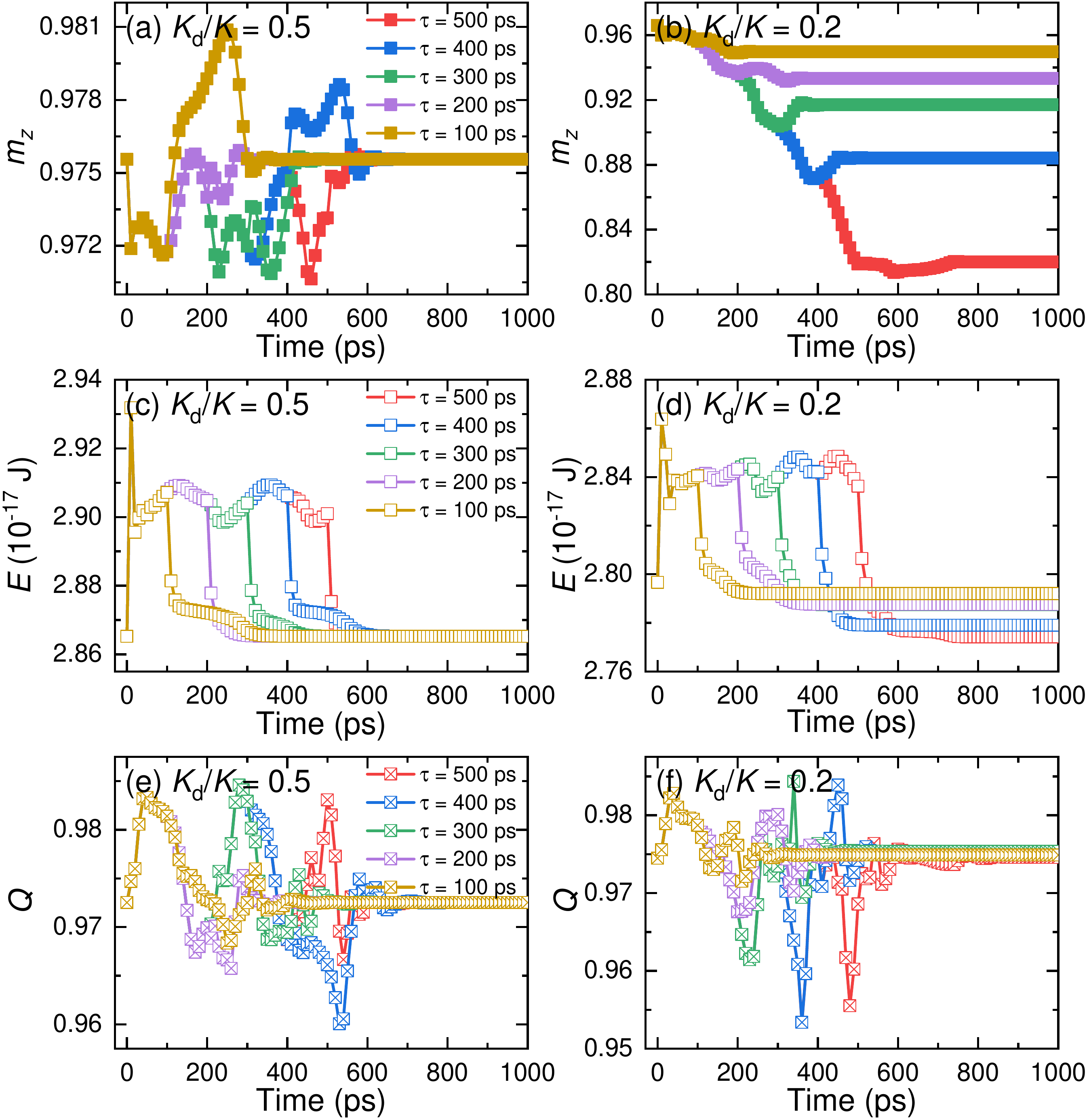}}
\caption{%
\textbf{Out-of-plane magnetization, energy, and topological charge of the system driven by current pulses with different pulse lengths.}
(\textbf{a}) Time-dependent out-of-plane magnetization $m_z$ corresponding to the current-induced hopping motion of a square-shaped skyrmion in a sample with $K_{\text{d}}/K=0.5$ for different pulse lengths $\tau$. $K$ and $K_{\text{d}}$ stand for the perpendicular magnetic anisotropy constants for the unmodified areas and defect lines, respectively. A current pulse of $j=100$ MA cm$^{-2}$ and $\boldsymbol{p}=-\hat{y}$ is applied, and then the system is relaxed until $t=1000$ ps.
(\textbf{b}) Time-dependent out-of-plane magnetization $m_z$ corresponding to the current-induced deformation of a square-shaped skyrmion in a sample with $K_{\text{d}}/K=0.2$ for different $\tau$.
(\textbf{c}) Time-dependent total energy $E$ of the system corresponding to (\textbf{a}).
(\textbf{d}) Time-dependent $E$ corresponding to (\textbf{b}).
(\textbf{e}) Time-dependent numerical topological charge $Q$ of the system corresponding to (\textbf{a}).
(\textbf{f}) Time-dependent numerical $Q$ corresponding to (\textbf{b}).
}
\label{FIG5}
\end{figure}

In Fig.~\ref{FIG5}(a), we show the time-evolution of $m_z$ corresponding to the hopping motion for various $\tau$ with $K_{\text{d}}/K=0.5$.
$m_z$ oscillates with time upon the pulse application and recovers to its initial value soon after the pulse application.
Such a feature of $m_z$ signal can be utilized to electrically detect the hopping motion of skyrmion.
$E$ increases and varies with time in a regular way during the pulse application [see Fig.~\ref{FIG5}(c)], implying the skyrmion moves regularly on the grid.
The variation of the numerical $Q$ is directly related to the skyrmion deformation [see Fig.~\ref{FIG5}(e)].
Next, we study the case with $K_{\text{d}}/K=0.2$.
Both $m_z$ and $E$ are decreased after the pulse application [see Fig.~\ref{FIG5}(b) and (d)], which indicates the rectangle-shaped and L-shaped skyrmions are stabler.
It is noteworthy that the topological nature of the skyrmion remains unchanged after its deformation [see Fig.~\ref{FIG5}(f)].
Since the area of a skyrmion on the grid is an integer multiple of the minimum grid cell area, the $E$ distribution of all possible pixelated skyrmions on the grid is discontinuous. 

\begin{figure}[t]
\centerline{\includegraphics[width=0.70\textwidth]{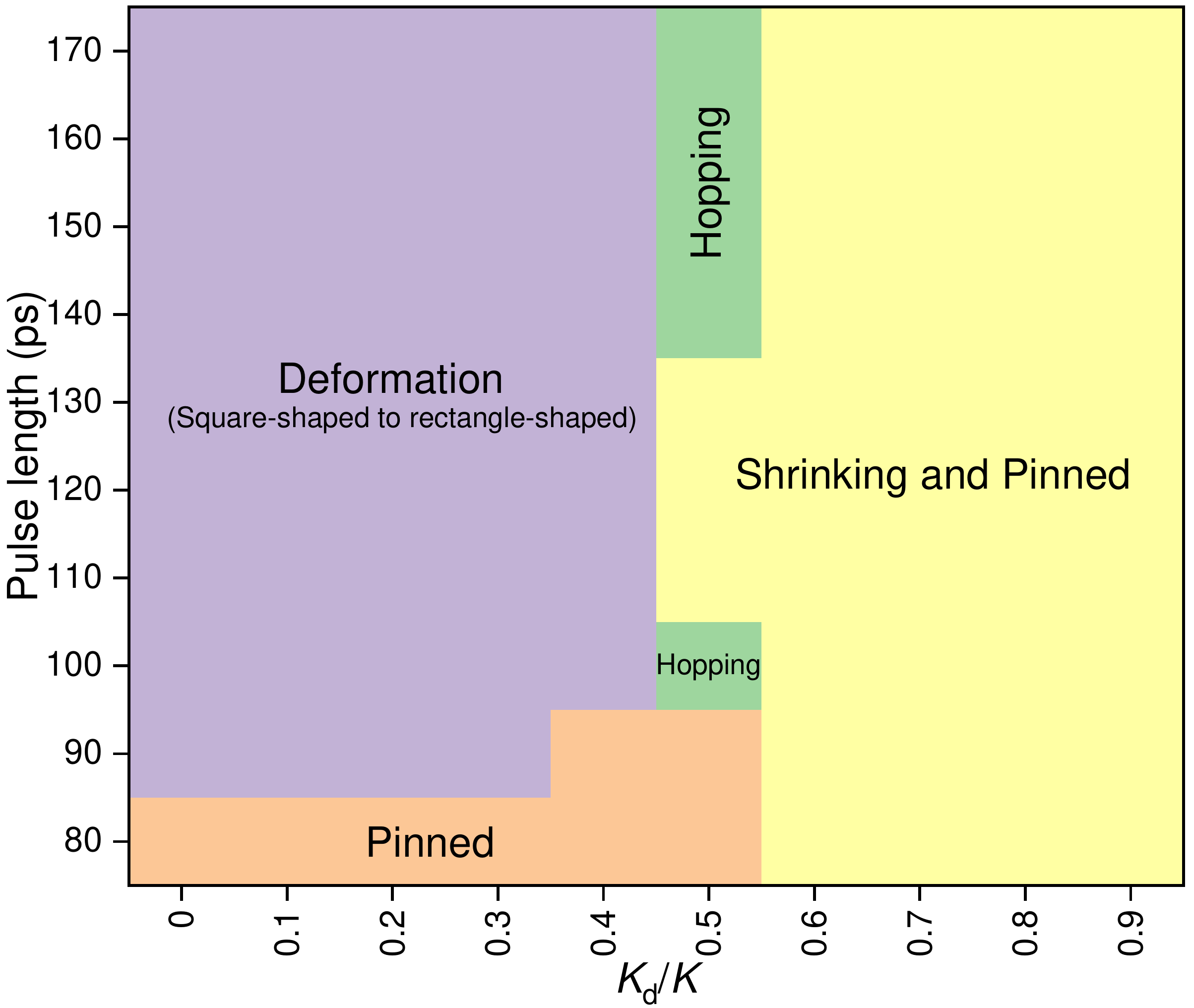}}
\caption{%
\textbf{A basic phase diagram of the system transitions from single skyrmion hopping to the skyrmion deformation.}
Deformation means the square-shaped skyrmion is deformed to a rectangle-shaped skyrmion after the pulse application. Hopping means the square-shaped skyrmion hops from the grid cell at the sample center to the right nearest-neighboring grid cell after the pulse application. Pinned means the square-shaped skyrmion is pinned by the grid cell at the sample center during and after the pulse application. Shrinking and pinned means the square-shaped skyrmion shrinks to a smaller round-shaped skyrmion and is pinned on the defect line after the pulse application. Here, $l=30$ nm, $w=4$ nm, $n=11$, and the damping parameter $\alpha=0.3$. $l$ denotes the spacing between two adjacent parallel defect lines. $w$ denotes the width of the defect line. $n$ denotes the number of unit square patterns along the $x$ and $y$ directions. The side length of the sample equals $378$ nm. A square-shaped skyrmion with the topological charge of $Q=1$ is relaxed at the sample center as the initial state at $t=0$ ps. A current pulse of $j=100$ MA cm$^{-2}$ and $\boldsymbol{p}=-\hat{y}$ is applied, and then the system is relaxed until $t=1000$ ps. The final states are confirmed at $t=1000$ ps.
}
\label{FIG6}
\end{figure}

We also study the effects of $j$ and $\alpha$ on the dynamics of a square-shaped skyrmion (see \blue{Supplementary Note 3}). The effect of $j$ is similar to that of $\tau$. A large $\alpha$ will reduce the skyrmion Hall effect for the skyrmion hop and deformation, which leads to the current-induced formation of the rectangle-shaped skyrmion.
A basic phase diagram of the system transitions from the single skyrmion hopping to the skyrmion deformation is given in Fig.~\ref{FIG6}.
We point out four possible cases induced by the current pulses with different pulse lengths. First, for the samples with relatively stronger pinning strengths ($K_{\text{d}}/K<0.5$), a weak current pulse cannot drive the square-shaped skyrmion. Namely, the square-shaped skyrmion is pinned at its initial position during and after the pulse application. For a strong current pulse, the square-shaped skyrmion will be transformed to a rectangle-shaped skyrmion by the current pulse. Second, for the sample with a moderate pinning strength ($K_{\text{d}}/K=0.5$), a weak current pulse cannot drive the square-shaped skyrmion, but a stronger current pulse may drive the square-shaped skyrmion into a hopping motion or shrinking. Third, for the sample with relatively weaker pinning strength ($K_{\text{d}}/K>0.5$), the square-shaped skyrmion may not be stable on the square pinning pattern. Hence, once a current pulse is applied, the square-shaped skyrmion will first depin and then shrink to a smaller round-shaped skyrmion. Such a smaller round-shaped skyrmion could be pinned again on the defect line after the pulse application.

%-%-%-%-%-%-%-%-%-%-%-%-%-%-%-%-%-%-%-%-%-%-%-%-%-%-%-%-%-%-%
\vbox{}
\section{\sffamily Conclusions}
%-%-%-%-%-%-%-%-%-%-%-%-%-%-%-%-%-%-%-%-%-%-%-%-%-%-%-%-%-%-%

\noindent
In conclusion, we have studied the statics and dynamics of configurable skyrmions on the grid formed by orthogonal defect lines with identical spacings and reduced PMA.
We find that the grid results in the pixelation of skyrmions, leading to the square-shaped, rectangle-shaped, and L-shaped skyrmions.
The position and area of the square-shaped skyrmion are quantized in the unit of the grid cell, which are different to ordinary skyrmions, of which the position and size change continuously.

We numerically demonstrate that the position, size, and shape of a square-shaped skyrmion on the grid are manipulated electrically, which depend on the pinning strength, the applied current pulse, and the damping parameter.
In particular, we show that the square-shaped skyrmion hops on the grid with weak pinning, and its skyrmion Hall effect can be controlled by the current pulse.
Especially, the skyrmion Hall effect of the square-shaped skyrmion is perfectly prohibited by appropriately tuning parameters.
The straight hopping motion of skyrmion is vital for racetrack-type memory devices.
The control of the skyrmion Hall effect using a preset sequence of current pulses provides the possibility to build a logic computing device based on the transport route of skyrmions.
In addition, it is possible to reduce the width of a nanotrack as wide as three grid cells since the skyrmion Hall effect is suppressed. It is highly contrasted with the case of an ordinary skyrmion, where we need to use a wider nanotrack in order to keep away a skyrmion from an edge.
It is possible to shift a skyrmion by $N$ grid cells by applying $N$ pulses since the skyrmion relaxes to the same structure only by changing its position after the pulse is over.

Besides, we find that the current pulse drives the square-shaped skyrmion to deform on the square grid with strong pinning, which transforms the square-shaped skyrmion to a rectangle-shaped or L-shaped skyrmion in a controlled manner.
It can be utilized to build a multistate memory~\cite{Wang_NC2020} or an artificial synapse~\cite{Song_NE2020} based on different metastable topological spin textures in one sample, where topological spin textures with different $m_z$ stand for different states that can be detected by measuring magnetoresistance.
It is worth mentioning that a reset function, that is, a method to transform an L-shaped skyrmion to an original square-shaped skyrmion may be required for the multistate memory and artificial synapse applications. Such a reset function can be achieved by applying an out-of-plane magnetic field pulse in our system (see \blue{Supplementary Note 4}).
Indeed, one can also reset the system by erasing the entire state and then nucleate a square-shaped skyrmion.

Our results give a deeper understanding of the complex dynamics of a skyrmion on a nanoscale grid formed by defect lines with modified magnetic anisotropy.
It will be straightforward to generalize our results to the systems with artificial nanoscale triangular and honeycomb grids.
However, the square-grid pinning pattern is most efficient to prohibit the skyrmion Hall effect and easily manufacturable, which is due to the fact that a typical lithography scanner system works in a way that favors horizontal and vertical scanning directions. For this reason, the fabrication of the triangular or irregular shape may result in obvious polygon edges. Such an effect may significantly reduce the pinning pattern quality when the resolution goes down to a few nanometers. Hence, the square-grid pinning pattern and rectangle-grid pinning pattern (see \blue{Supplementary Note 5}) may be the most reliable choices.
Besides, the advantage of using the square-grid pinning pattern to guide the skyrmion motion is that the skyrmion can be delivered toward different directions by controlling the driving current direction, current density, and pulse length. Such a feature may not be possible on other pinning patterns such as the parallel defect lines.

On the other hand, we would like to point out that the square-grid pinning pattern could also serve as a platform for the study of multiple skyrmions interacting with a pinning landscape (see \blue{Supplementary Note 6}), and a great many directions one could go with this system such as different kinds of driving forces~\cite{Reichhardt_2021}.
Last, from the point of view of electronic device applications, future works on this topic may focus on the performance analysis, such as the energy expenditures of skyrmion hopping and square-to-L deformation.
Our results may provide guidelines for building spintronic applications utilizing the interaction between topological spin textures and artificial pinning patterns.

%-%-%-%-%-%-%-%-%-%-%-%-%-%-%-%-%-%-%-%-%-%-%-%-%-%-%-%-%-%-%
\vbox{}
\section{\sffamily Methods}
%-%-%-%-%-%-%-%-%-%-%-%-%-%-%-%-%-%-%-%-%-%-%-%-%-%-%-%-%-%-%

\noindent
\textbf{Micromagnetic simulations.}
All spin relaxation and dynamics simulations are carried out by using the Object Oriented MicroMagnetic Framework (OOMMF) developed at NIST~\cite{OOMMF}.
The three-dimensional spin dynamics in the FM sample is governed by the Landau-Lifshitz-Gilbert (LLG) equation augmented with the damping-like spin-orbit torque~\cite{OOMMF}
\begin{equation}
\begin{split}
\label{eq:LLG-SHE}
\frac{d\boldsymbol{M}}{dt}=&-\gamma_{0}\boldsymbol{M}\times\boldsymbol{H}_{\text{eff}}+\frac{\alpha}{M_{\text{S}}}(\boldsymbol{M}\times\frac{d\boldsymbol{M}}{dt}) \\
&+\frac{u}{M_{\text{S}}}(\boldsymbol{M}\times \boldsymbol{p}\times \boldsymbol{M}),
\end{split}
\end{equation}
where the damping-like spin torque is generated through the spin Hall effect in the heavy-metal layer when an electric current is injected~\cite{Tomasello_SREP2014}.
In Eq.~\ref{eq:LLG-SHE}, $\boldsymbol{M}$ is the magnetization, $M_{\text{S}}=|\boldsymbol{M}|$ is the saturation magnetization, $t$ is the time, $\gamma_{\text{0}}$ is the absolute value of gyromagnetic ratio, $\alpha$ is the Gilbert damping parameter, and $\boldsymbol{H}_{\text{eff}}=-\mu_{0}^{-1} \partial\varepsilon/\partial \boldsymbol{M}$ is the effective field.
$u=|(\gamma_{0}\hbar)/(\mu_{0}e)|\cdot(j\theta_{\text{SH}})/(2aM_{\text{S}})$ is the spin torque coefficient, $\boldsymbol{p}$ stands for the unit spin polarization direction, $\mu_0$ is the vacuum permeability constant, $\hbar$ is the reduced Planck constant, $e$ is the electron charge, $j$ is the driving current density, and $\theta_{\text{SH}}$ is the spin Hall angle.

The average energy density $\varepsilon$ contains the PMA, FM exchange, demagnetization, applied magnetic field, and interface-induced DM interaction energy terms, given as
\begin{equation}
\label{eq:energy-density} 
\begin{split}
\varepsilon=&-K\frac{(\boldsymbol{n}\cdot\boldsymbol{M})^{2}}{M_{\text{S}}^{2}}+A\left[\nabla\left(\frac{\boldsymbol{M}}{M_{\text{S}}}\right)\right]^{2}-\frac{\mu_{0}}{2}\boldsymbol{M}\cdot\boldsymbol{H}_{\text{d}} \\
&-\boldsymbol{M}\cdot\boldsymbol{B}+\frac{D}{M_{\text{S}}^{2}}\left[M_{z}\left(\boldsymbol{M}\cdot\nabla\right)-\left(\nabla\cdot\boldsymbol{M}\right)M_{z}\right],
\end{split}
\end{equation}
where $K$, $A$, and $D$ are the PMA, FM exchange, and DM interaction energy constants, respectively. $\boldsymbol{B}$ is the applied magnetic field, and $\boldsymbol{H}_{\text{d}}$ is the demagnetization field. $\boldsymbol{n}$ is the unit surface normal vector. $M_z$ is the out-of-plane component of $\boldsymbol{M}$. 
The default parameters used in this work are~\cite{Sampaio_NN2013,Tomasello_SREP2014,Xichao_NC2016}: $\gamma_{0}=2.211\times 10^{5}$ m A$^{-1}$ s$^{-1}$, $\alpha=0.05\sim 0.5$, $M_{\text{S}}=580$ kA m$^{-1}$, $K=0.8$ MJ m$^{-3}$, $A=15$ pJ m$^{-1}$, $D=3$ mJ m$^{-2}$, and $\theta_{\text{SH}}=0.2$. The mesh size is set as $1$ $\times$ $1$ $\times$ $1$ nm$^3$ in all simulations, guaranteeing both accuracy and efficiency.

\vbox{}
\noindent
\textbf{Data availability.}
The data that support the findings of this study are available from the corresponding authors upon reasonable request.

\vbox{}
\noindent
\textbf{Code availability.}
The micromagnetic simulator OOMMF used in this work is publicly accessible at http://math.nist.gov/oommf.

\vbox{}
\clearpage

%-%-%-%-%-%-%-%-%-%-%-%-%-%-%-%-%-%-%-%-%-%-%-%-%-%-%-%-%-%-%

%-%-%-%-%-%-%-%-%-%-%-%-%-%-%-%-%-%-%-%-%-%-%-%-%-%-%-%-%-%-%

%-%-%-%-%-%-%-%-%-%-%-%-%-%-%-%-%-%-%-%-%-%-%-%-%-%-%-%-%-%-%
\vbox{}
\section{\sffamily Acknowledgements}
\noindent
%-%-%-%-%-%-%-%-%-%-%-%-%-%-%-%-%-%-%-%-%-%-%-%-%-%-%-%-%-%-%
%
X.Z. was an International Research Fellow of the Japan Society for the Promotion of Science (JSPS). X.Z. was supported by JSPS KAKENHI (Grant No. JP20F20363). O.A.T. acknowledges the support by the Australian Research Council (Grant No. DP200101027), NCMAS grant, and the Cooperative Research Project Program at the Research Institute of Electrical Communication, Tohoku University. M.E. acknowledges the support by the Grants-in-Aid for Scientific Research from JSPS KAKENHI (Grant Nos. JP17K05490 and JP18H03676) and the support by CREST, JST (Grant Nos. JPMJCR16F1 and JPMJCR20T2). Y.Z. acknowledges the support by the Guangdong Special Support Project (Grant No. 2019BT02X030), Shenzhen Fundamental Research Fund (Grant No. JCYJ20210324120213037), Shenzhen Peacock Group Plan (Grant No. KQTD20180413181702403), Pearl River Recruitment Program of Talents (Grant No. 2017GC010293), and National Natural Science Foundation of China (Grant Nos. 11974298 and 61961136006). X.L. acknowledges the support by the Grants-in-Aid for Scientific Research from JSPS KAKENHI (Grant Nos. JP20F20363, JP21H01364, and JP21K18872).

%-%-%-%-%-%-%-%-%-%-%-%-%-%-%-%-%-%-%-%-%-%-%-%-%-%-%-%-%-%-%
\vbox{}
\section{\sffamily Author contributions}
%-%-%-%-%-%-%-%-%-%-%-%-%-%-%-%-%-%-%-%-%-%-%-%-%-%-%-%-%-%-%

\noindent
X.L. and X.Z. conceived the idea. X.L., M.E. and Y.Z. coordinated the project. X.Z. and J.X. performed the computational simulations and data analysis. M.E. and O.A.T. carried out the theoretical analysis. X.Z. drafted the manuscript and revised it with input from J.X., K.S., H.F., M.E., O.A.T., Y.Z. and X.L. All authors discussed the results and reviewed the manuscript. X.Z. and J.X. contributed equally to this work.

%-%-%-%-%-%-%-%-%-%-%-%-%-%-%-%-%-%-%-%-%-%-%-%-%-%-%-%-%-%-%
\vbox{}
\section{\sffamily Competing interests}
%-%-%-%-%-%-%-%-%-%-%-%-%-%-%-%-%-%-%-%-%-%-%-%-%-%-%-%-%-%-%

\noindent
The authors declare no competing interests.

%-%-%-%-%-%-%-%-%-%-%-%-%-%-%-%-%-%-%-%-%-%-%-%-%-%-%-%-%-%-%
\vbox{}
\section{\sffamily Additional information}
%-%-%-%-%-%-%-%-%-%-%-%-%-%-%-%-%-%-%-%-%-%-%-%-%-%-%-%-%-%-%

\noindent
\textbf{Supplementary information} The online version contains supplementary material available at [\href{https://www.nature.com/articles/s42005-021-00761-7}{\blue{https://www.nature.com/articles/s42005-021-00761-7}}].

\vbox{}
\noindent
\textbf{Correspondence} and requests for materials should be addressed to M.E., Y.Z. or X.L.

\clearpage
\noindent
\textbf{\sffamily Figures and Captions}

%%%%%%%%%%%%%%%%%%%%%%%%%%%%%%%%%%%%%%%%%%%%%%%%%%%%%%%%%%%%
%%%%%%%%%%%%%%%%%%%%%%%%%%%%%%%%%%%%%%%%%%%%%%%%%%%%%%%%%%%%

%%%%%%%%%%%%%%%%%%%%%%%%%%%%%%%%%%%%%%%%%%%%%%%%%%%%%%%%%%%%
%%%%%%%%%%%%%%%%%%%%%%%%%%%%%%%%%%%%%%%%%%%%%%%%%%%%%%%%%%%%

%%%%%%%%%%%%%%%%%%%%%%%%%%%%%%%%%%%%%%%%%%%%%%%%%%%%%%%%%%%%
%%%%%%%%%%%%%%%%%%%%%%%%%%%%%%%%%%%%%%%%%%%%%%%%%%%%%%%%%%%%

%%%%%%%%%%%%%%%%%%%%%%%%%%%%%%%%%%%%%%%%%%%%%%%%%%%%%%%%%%%%
%%%%%%%%%%%%%%%%%%%%%%%%%%%%%%%%%%%%%%%%%%%%%%%%%%%%%%%%%%%%

%%%%%%%%%%%%%%%%%%%%%%%%%%%%%%%%%%%%%%%%%%%%%%%%%%%%%%%%%%%%
%%%%%%%%%%%%%%%%%%%%%%%%%%%%%%%%%%%%%%%%%%%%%%%%%%%%%%%%%%%%

%%%%%%%%%%%%%%%%%%%%%%%%%%%%%%%%%%%%%%%%%%%%%%%%%%%%%%%%%%%%
%%%%%%%%%%%%%%%%%%%%%%%%%%%%%%%%%%%%%%%%%%%%%%%%%%%%%%%%%%%%

%%%%%%%%%%%%%%%%%%%%%%%%%%%%%%%%%%%%%%%%%%%%%%%%%%%%%%%%%%%%
\end{document}